**Crystal Nucleation and Growth in Liquids: Cooperative Atom Attachment and Detachment**

Fangzheng Chen,[1] Zohar Nussinov[2,3] and K. F. Kelton[2,1]

1. Institute of Materials Science and Engineering, Washington University in St. Louis, St. Louis, Missouri 63130 USA
2. Department of Physics, Washington University in St. Louis, St. Louis, Missouri 63130 USA
3. Rudolf Peierls Centre for Theoretical Physics, University of Oxford, Oxford 0X1 3PU, United Kingdom

**Abstract**

Classical theories of crystal nucleation and growth from the liquid assume activated processes that are interface limited, with the atoms individually joining the growing interface by jumps that occur at a rate that is determined by the diffusion coefficient in the liquid phase. These assumptions are in contradiction with the results of molecular dynamics studies that are presented here for supercooled Ni and $Al_{20}Ni_{60}Zr_{20}$. Instead of diffusion-based attachment across the interface, atoms join the interface by making small changes so as to match the orientational order parameter of the nucleating crystal. Further, instead of joining individually multiple atoms join cooperatively, with the number of cooperative atoms increasing with decreasing temperature.

Crystallization in liquids or glasses consists of a nucleation step in which clusters spontaneously grow and shrink stochastically. When a cluster becomes sufficiently large (exceeding the critical size of $n^*$ atoms) it is biased to continue to grow. Two fundamental assumptions are made in the widely used Classical Nucleation Theory (CNT). First, the interface between the nucleating cluster and the original phase is sharp and second, each step in the cluster development is governed by individual atoms attaching or detaching from the cluster interface. While these assumptions are valid for gas condensation, the process that the CNT was originally developed to describe, they are questionable in the Turnbull adaptation of CNT to describe crystal nucleation from a liquid [1]. The structure of the liquid that is adjacent to the nucleating phase is characterized by short- and medium-range order that may even be similar to that of the nucleating ordered phase; because of this the interface is not sharp. This has been confirmed by experimental nucleation studies, density functional calculations and molecular dynamics simulations (see chapters 4, 7 and 10 in ref [2]). In contrast there has been very little investigation of the assumed path for atom attachment to the cluster interface, *i.e.,* whether atoms join singly as assumed in CNT or whether several atoms might join cooperatively. A recently developed analytical model for crystal growth assumed the latter [3, 4] and used an Adam-Gibbs approach [5] to model the cooperative attachment.

In this letter, results from a molecular dynamics (MD) simulation study of crystal nucleation and growth in two metallic supercooled liquids, Ni and $Al_{20}Ni_{60}Zr_{20}$, are presented that are in conflict with some of the assumptions of CNT. For both liquids the MD results show that rather than acting as single atoms, multiple nearest-neighbor atoms attach and detach cooperatively. Further, the number of atoms acting cooperatively (a measure of the coherence length) increases with decreasing temperature. Finally, the atoms join the interface by subtle changes in their order parameter and do not involve the significant movement in space assumed in the classical theories of nucleation and growth.

The MD simulations were made using the Large-scale Atomic/Molecular Massively Parallel Simulator (LAMMPS) in the Extreme Science and Engineering Discovery Environment (XSEDE) [6]. Embedded atom (EAM) potentials for $Al_{20}Ni_{60}Zr_{20}$ [7] and Ni [8] were used to describe the atomic interactions. The ensemble for the $Al_{20}Ni_{60}Zr_{20}$ liquid was created by randomly assigning 5000 Al atoms, 5000 Zr atoms and 15000 Ni atoms at 2500 K; this was relaxed for 2 nanoseconds (ns) to reach equilibrium. The ensemble was then cooled to the target temperature at a constant cooling rate of 10 K per picosecond (10 K/ps) and equilibrated again. The monoatomic Ni liquid, which consisted of 32,000 atoms, was created following the same procedure as for the $Al_{20}Ni_{60}Zr_{20}$ liquid. Studies of nucleation and growth were made in two ways, one by inserting a crystal seed into the supercooled liquid and a second by waiting for clusters to form spontaneously in the liquid (i.e., homogeneous nucleation). All of the simulations were made using the NPT (isobaric-isothermal) ensemble with periodic boundary conditions.

For the $Al_{20}Ni_{60}Zr_{20}$ liquid, the critical size for nucleation, $n^*$, was determined at 1150 K using the seeding method, *i.e.* inserting clusters of different size in the liquid and observing whether they shrink or grow [9]. The probability to shrink or grow will be equal for a cluster of size $n^*$. After equilibrating each seed for 5 ps to heal the interface with the liquid, the system was equilibrated for another 5 ps. The couple (seed and liquid) was then annealed for one nanosecond and the final number of atoms in the seed was recorded. By this method, $n^*$ at 1150 K was found to contain 366 atoms. A cluster containing 371 atoms (larger than the critical size) was then inserted into the $Al_{20}Ni_{60}Zr_{20}$ metallic liquid at 1150 K and cluster development was monitored.

To better identify the liquid/crystal atoms, an index of crystallinity (IC) was constructed from the local bond-order parameter $q_6$ [10, 11]. The quantity

$$\vec{q}_6(i) \cdot \vec{q}_6(j) = \sum_{m=-6}^{m=6} \tilde{q}_{6m}(i) \tilde{q}_{6m}(j)^* \qquad (1)$$

(where $\tilde{q}_{6m}$ is the normalized local orientational order parameter) measures how similar the environments are for neighboring atoms *i* and *j*. They are considered to be connected if the dot product exceeds a certain threshold, which was 0.3 in this study. A cross section of a typical nucleating cluster is shown in fig. 1. The IC values are color coded, ranging from approximately 120 in the center of the cluster (red) to about 50 for atoms at the cluster/liquid interface. The average IC values for atoms in the liquid (not shown) are smaller than 40. This shows that the order parameter increases on moving to the center of the cluster. It also shows that while not perfectly spherical the cluster is reasonably compact, as expected from CNT, although the cluster/liquid boundary is diffuse rather than sharp with an interfacial width of two to three atomic layers. This is in agreement with density functional calculations and is assumed in the Diffuse Interface Theory (see chapter 4 in ref. [2]). However, two points should be noted: (1) this cluster is larger than $n^*$ at this temperature (1150 K) and (2) the inserted cluster is spherical, so the compactness might be a remnant of that. Smaller clusters are less compact and the interface is more ragged.

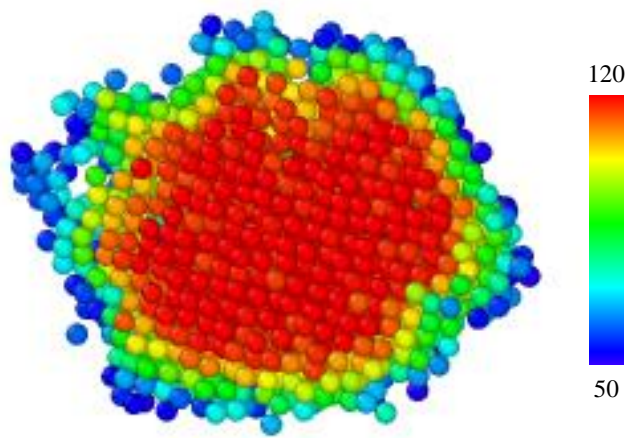

**Figure 1** – A cross section of a growing cluster, showing color coding for the atom IC values. The crystalline order in the center of the cluster is evident.

The upper left inset in fig. 2 shows the change in IC with time for a typical atom (ID 15867) that was initially in the liquid (at $t = 0$), but begins to incorporate into the interface of the crystal phase (at ≈ 450 ps) and finally takes on the stable value of the crystal (at ≈ 600 ps). The attachment behavior of the nearest neighbor atoms to the target atom is also shown. Nearest neighbors are defined as atoms that are within 3.5 Å, which is the first minimum after the first peak in the calculated $g(r)$, for at least 80% of the time. Remarkably, the time dependence of the ICs for the neighboring atoms during cluster attachment track those of the target atom and of each other. The width of the rise is approximately 100 ps for each atom, reflecting the stochastic nature of nucleation. The highly overlapped curves suggest that the atoms are attaching cooperatively, *i.e.,* as a group of atoms, rather than the single atom attachment model assumed in CNT.

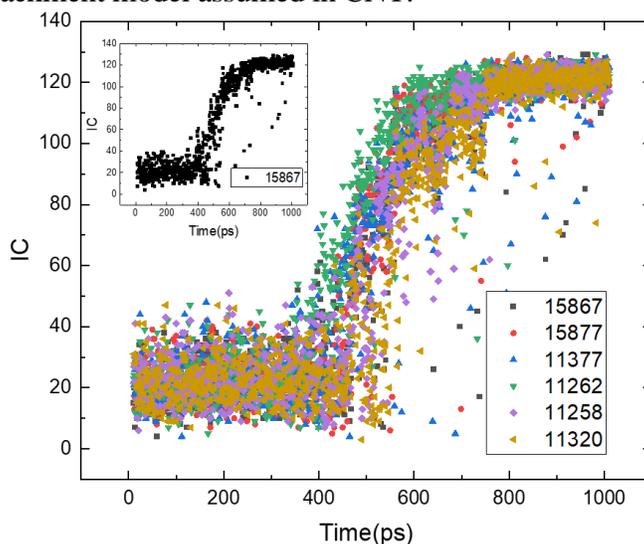

**Figure 2** – IC as a function of time. The inset shows this for the target atom (atom ID 15867). The main figure shows how this correlates with the IC as a function of time for the target atom and its nearest neighbor atoms.

To better visualize the collective motion, the positions of the six atoms discussed in fig. 2 are shown in fig 3. Figure 3.a shows the atoms in the liquid at the cluster interface prior to attaching to the cluster

(blue indicating a low IC value). After 480 ps their IC value has increased to near 50 as the atoms move a small distance (much less than the atomic-distance jumps envisioned in the classical theories of nucleation and growth) to attach to the cluster (fig. 3.b). To better observe how the atoms are incorporated into the cluster, the color of this group of atoms is changed to white (fig. 3.c) (note that this does not correspond to a particular IC value). As shown in fig. 3.d, these atoms are incorporated into the cluster interior as other atoms attach to the interface. It is important to emphasize again that these atoms do not move a significant distance to attach to the cluster interface, as assumed within the classical theories of nucleation and growth, but instead join by the evolution of their order parameter towards that of the crystal cluster (by small changes in orientation to the nucleating interface). To further clarify this, the average distance moved during each ps for atoms with different IC values are listed in Table S1 in the supplemental section. A similar attachment behavior was observed for clusters much larger than the critical size, supporting the model for growth that was recently proposed [3].

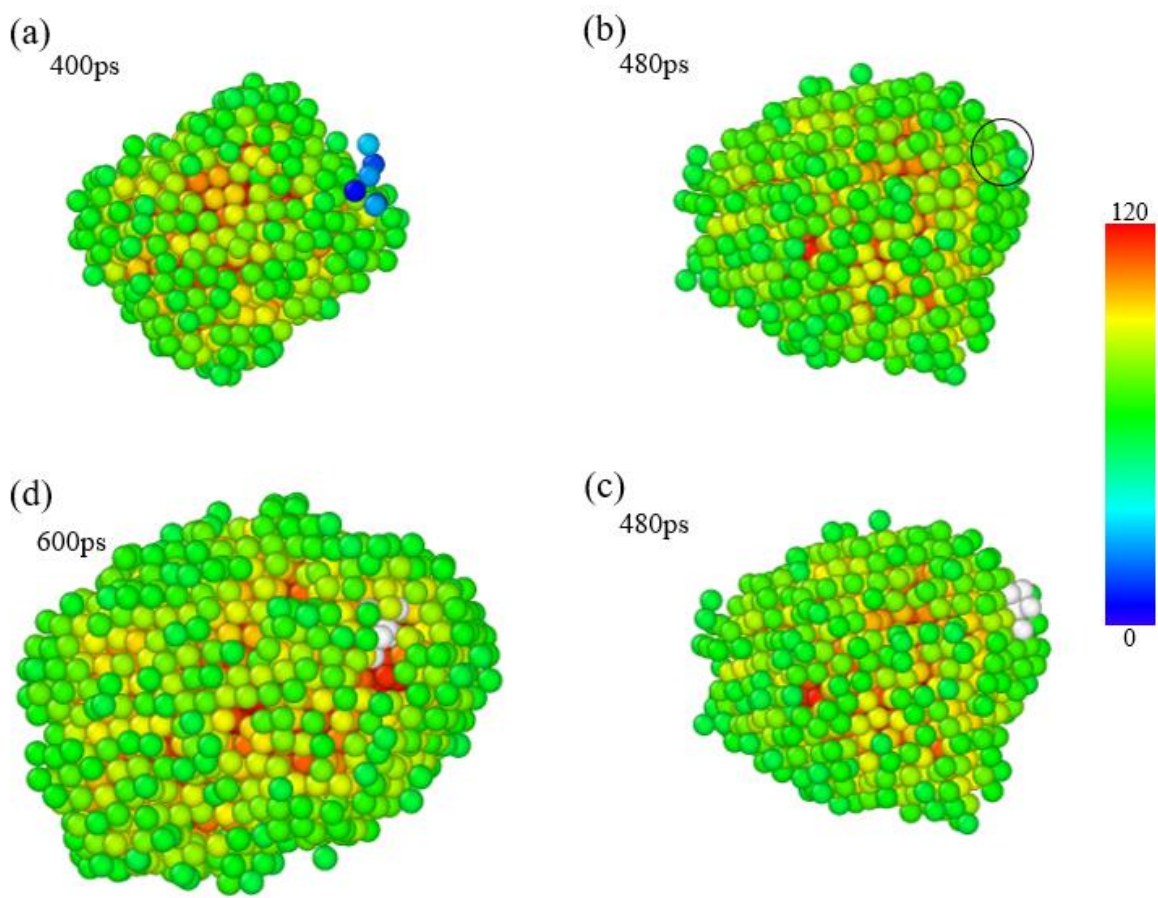

**Figure 3** – The collective motion of a group of atoms. (a) A group of liquid atoms (IC close to 0) are at the interface of the cluster before attachment. (b) Atoms that attach to the cluster (within the black circle) via collective motion. Their IC value increases to around 50 at 480 ps. (c) The color of the attaching atoms changed to white for better visualization. Note that the color white does not represent an IC value. (d) The targets atoms incorporated into the cluster at 600 ps as other liquid atoms attach to the cluster interface.

So far, the MD analysis has focused on atom attachment, which is important in both nucleation and growth. But nucleation is a stochastic process, with atoms joining and leaving the cluster at similar rates. A natural question is whether detachment is also a collective process. To examine this, the dissolution of an inserted critical cluster at 1150 K was followed. A target atom (ID 24711) was randomly selected from the cluster and its IC was collected over 1 ns. As shown in the inset to fig. 4, this atom left the crystal to join the liquid over a time interval from 100 ps to 200 ps. Using the same procedure already discussed for attachment, the detachment of the nearest neighbor atoms to the target atom was tracked as a function of time. As shown in fig. 4 the detachment behavior of all of the neighboring atoms tracks that of the target atom and of each other. There is even less fluctuation among the group of atoms than for the case of attachment. These results indicate that the collective behavior exists for both attachment to and detachment from the nucleating cluster.

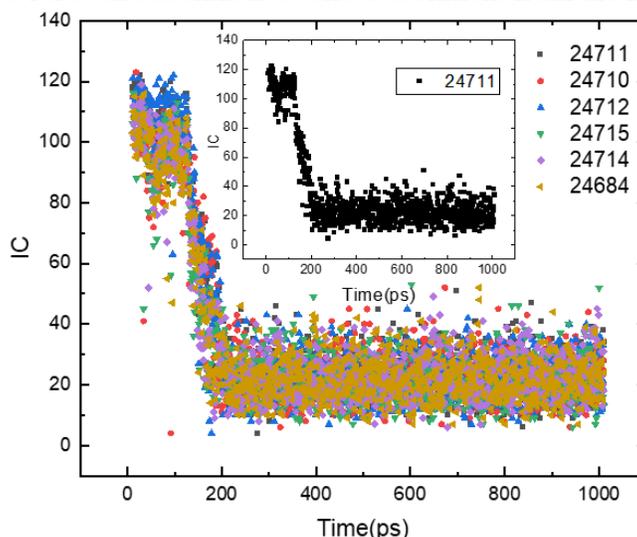

**Figure 4** – (inset) The IC value for the target atom (ID 24711) as a function of time during detachment. The IC values for the nearest neighbors to the target atom.

It might be argued that the collective behavior is an artifact of the seeding procedure. To check this homogenous nucleation was studied. An $Al_{20}Ni_{60}Zr_{20}$ liquid ensemble of 25,000 atoms was held at 1050 K for 2 ns and a target atom that eventually joined the nucleating cluster (ID 22546) was followed as a function of time. As shown in fig. 5.a, the target atom initially joined and then detached from one nucleating cluster during the time interval of 650 ps to 850 ps; it eventually attached to another nucleating cluster after 1.8 ns. In fig 5.b the nearest neighbor atoms changed their IC value cooperatively with the target atom during detachment from the first nucleating cluster (fig. 5.b). After 850 ps, these atoms joined the liquid again and diffused away from each other. As it moved towards the interface of the second nucleating cluster, the target atom gained new neighbors. When it attached to the new cluster, these neighbors joined cooperatively with the target atom (fig. 5.c). This clearly shows that attachment and detachment during nucleation as well as growth is in general cooperative, which conflicts with the classical theories of nucleation and growth. Further investigations showed the same behavior for nucleation in liquid Ni (discussed in the supplementary materials section). A recent model for crystal growth assumed that the cooperativity could be described with the Adam-Gibbs model [3]. This suggests that the coherence length for cooperativity (or equivalently the number of coherent atoms) should increase with decreasing temperature, which is the case as shown in Table S.2 in the supplemental section. The coherence length is nearly the same for all atoms, irrespective of their elemental identity.

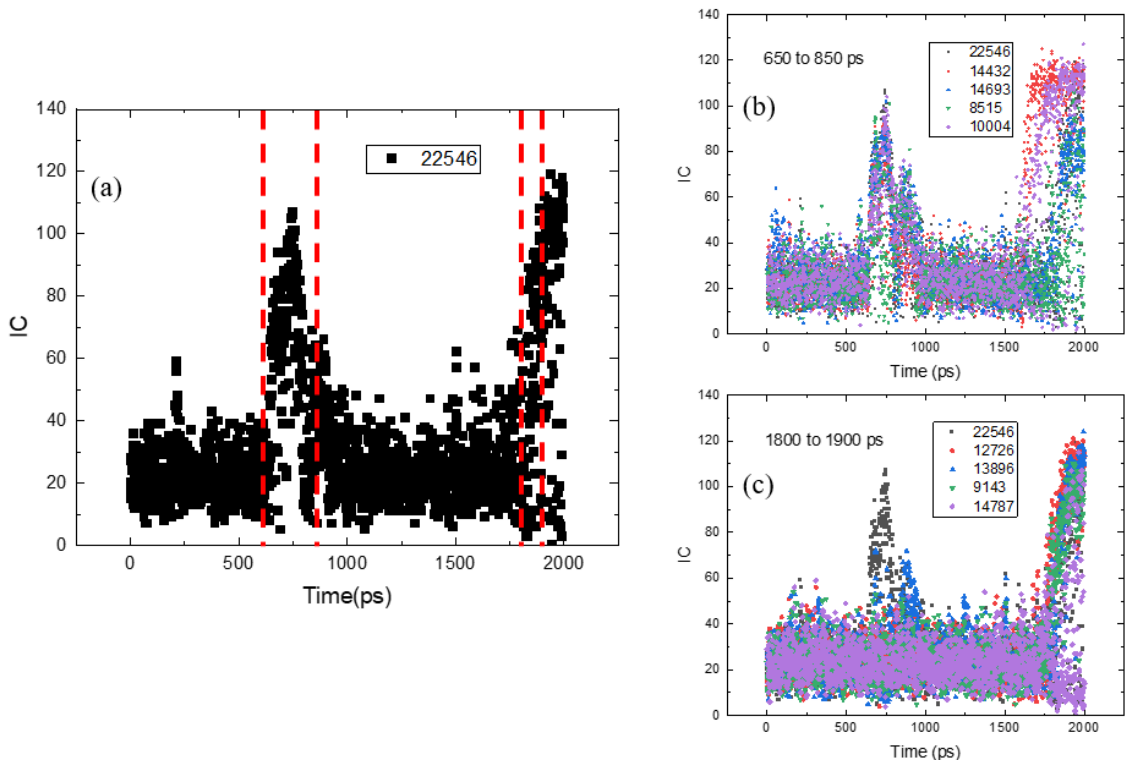

**Figure 5** – Cooperative motion of a single atom (atom ID 22546) for homogeneous nucleation. The dashed red lines indicate two attachment events. (b) A sub-cluster was formed with the target atom, but then dissolved from 650 ps to 850 ps. The nearest neighbor atoms show a highly overlaped IC plot with the target atom during this time. However, they diffused away from each other after cluster dissolution and have different IC values at the end. (c) The target atom attached to a growing cluster in the time interval between 1800 ps and 1900 ps. It has a similar IC change with its neighbor atoms at the end of the simulation.

It should be pointed out that cooperativity has also been identified from MD studies of the attachment of Al during crystal nucleation in an Al-Sm glass [12]. The detachment of Al in nucleation was not considered, however, nor was the incorporation of the Sm atoms into the interface. The study was made only in glasses with low Sm concentration and it was suggested that collective behavior might not be true for growth or crystal nucleation in alloys containing larger concentrations of Sm. Our studies of nucleation and growth in $Al_{20}Ni_{60}Zr_{20}$ and Ni metallic liquids, however, suggest that cooperative attachment/detachment is a universal property.

In conclusion, the nucleation of $Al_{20}Ni_{60}Zr_{20}$ and monoatomic Ni metallic liquids were studied in classical MD simulations. These studies show that nucleation does not require atom motion over atomic lengths as assumed in the Classical Theory of Nucleation (CNT). Also, rather than the single atom attachment/detachment assumed in the kinetic model for CNT, a small group of nearest neighbor atoms in the liquid near the interface attach and detach to the cluster by cooperatively following small changes in their order parameter. No evidence for cooperative behavior was observed for atoms in the liquid.  To our knowledge, this is the first quantitative and comprehensive evidence for the collective motion of atoms for both heterogeneous and homogenous nucleation.  That it occurs in both metallic liquids and glasses suggests that it a pervasive kinetic process that should be taken into account for the refinement of nucleation and growth theories.  The extent to which it influences physical nucleation processes should also be investigated.

## Acknowledgements

We thank A. L. Greer for helpful discussion.  The research was partially supported by the National Science Foundation under Grant No. DMR-19-04281.  Any opinions, findings, and conclusions or recommendations expressed in this material are those of the author(s) and do not necessarily reflect the views of the National Science Foundation.

## Data Availability

The data that support the findings of this study are available from the corresponding author upon request.